\documentclass[prb,aps,superscriptaddress,twocolumn,showkeys,floatfix]{revtex4-1}

\usepackage{graphicx}
\usepackage{hyperref}

\begin{document}

\title{Hydrogen-induced reversible spin-reorientation transition and magnetic stripe domain phase in bilayer Co on Ru(0001)}

\author{Benito Santos}
\altaffiliation[Present address: ]{Elettra Sincrotrone S.C.p.A, 34149 Basovizza, Trieste, Italy}
\affiliation{Instituto de Qu\'{\i}mica-F\'{\i}sica ``Rocasolano'', CSIC, Madrid 28006, Spain}
\author{Silvia Gallego}
\affiliation{Instituto de Ciencia de Materiales de Madrilsd, CSIC, Madrid 28049, Spain}
\author{Arantzazu Mascaraque}
\affiliation{Dpto. de F\'{\i}sica de Materiales (Unidad Asociada IQFR-CSIC), Universidad Complutense de Madrid, Madrid 28040, Spain}
\affiliation{Unidad Asociada IQFR(CSIC)-UCM, Madrid 28040, Spain}
\author{Kevin F. McCarty}
\affiliation{Sandia National Laboratories, Livermore, California 94550, USA}
\author{Adrian Quesada}
\altaffiliation[Present address: ]{Instituto de Cer\'{a}mica y Vidrio, CSIC, Madrid 28049, Spain}
\affiliation{Lawrence Berkeley National Laboratory, Berkeley, California 94720, USA}
\author{Alpha T. N'Diaye}
\affiliation{Lawrence Berkeley National Laboratory, Berkeley, California 94720, USA}
\author{Andreas K. Schmid}
\affiliation{Lawrence Berkeley National Laboratory, Berkeley, California 94720, USA}
\author{Juan de la Figuera}
\affiliation{Instituto de Qu\'{\i}mica-F\'{\i}sica ``Rocasolano'', CSIC, Madrid 28006, Spain}
\email{juan.delafiguera@iqfr.csic.es}

\begin{abstract}
Imaging the change in the magnetization vector in real time by spin-polarized low-energy electron microscopy, we observed a hydrogen-induced, reversible spin-reorientation transition in a cobalt bilayer on Ru(0001). Initially, hydrogen sorption reduces the size of out-of-plane magnetic domains and leads to the formation of a magnetic stripe domain pattern, which can be understood as a consequence of reducing the out-of-plane magnetic anisotropy. Further hydrogen sorption induces a transition to an in-plane easy-axis. Desorbing the hydrogen by heating the film to 400~K recovers the original out-of-plane magnetization. By means of ab-initio calculations we determine that the origin of the transition is the local effect of the hybridization of the hydrogen orbital and the orbitals of the Co atoms bonded to the absorbed hydrogen.
\end{abstract}

\date{\today}


\maketitle 

\section{Introduction}

Gas adsorption\cite{somorjai_introduction_2010} can modify key properties of ultrathin films, such as the structure and atomic layer spacings of metallic films or surfaces. Gas adsorption can also significantly affect magnetic properties of ultrathin films. Even relatively subtle interactions, such as charge transfer between adsorbates and the metal atoms, can introduce the modification of the magnetic moment of the topmost atoms of the metal film. Already nearly fifty years ago changes in the magnetization of ferromagnetic materials produced by the chemisorption of hydrogen were reported\cite{abeledo_chemisorption_1962}. Since then, many studies have found that adsorption of gases can have important effects on surface magnetism, such as inducing either a decrease or an increase in the magnetic moment of the topmost atoms of a ferromagnetic film; several excellent reviews\cite{JohnsonReview,VazBlandLauhoffReview} discuss many interesting examples.

\begin{figure*}[tb]
\centering \includegraphics[width=1.0\textwidth]{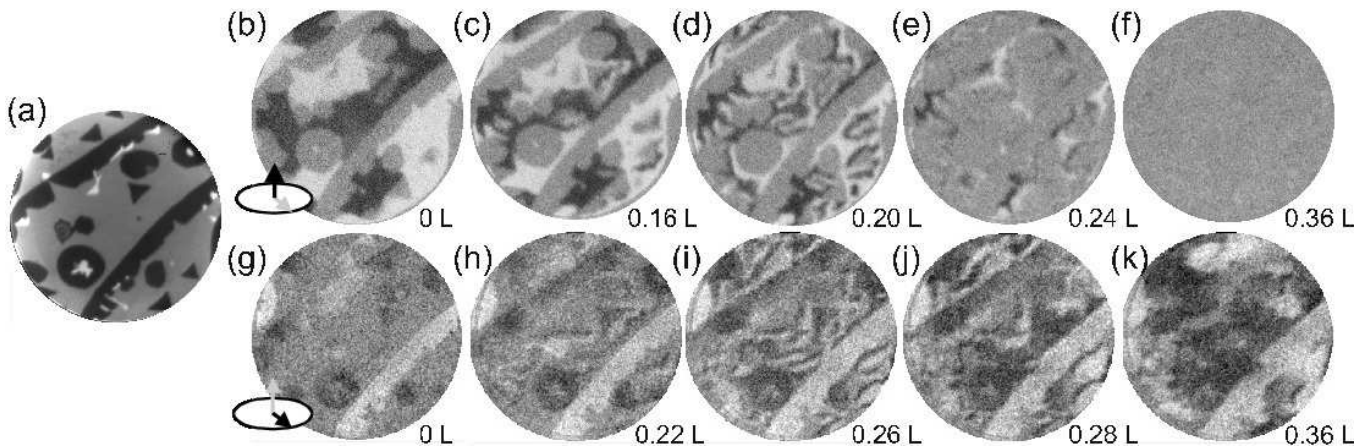}
\caption{(a) LEEM image of a Co film with 3~ML Co islands (dark areas) on a nearly continuous 2~ML film (medium gray areas). (b)--(f) SPLEEM images with out-of-plane magnetic domain contrast as a function of hydrogen exposure.  (g)--(k) SPLEEM images showing the in-plane magnetic domains of the same region during hydrogen exposure after desorbing the initial hydrogen by annealing to 400 K. The dose is indicated in the lower right corner of each SPLEEM image. Field of view (FOV) is 4 $\mu$m, beam energy is 5.2 eV.}
\label{durante}
\end{figure*} 
 
Besides affecting the value of the magnetic moment, gas sorption can also change the magnetic anisotropy of surfaces and films. Hydrogen has been observed to induce a spin-reorientation transition (SRT) in Ni/Cu films. In that case, hydrogen-induced strain effects have been invoked to explain the SRT\cite{sander_reversible_2004}. Residual gas absorption on Fe/W(110) films has been observed to result in an SRT\cite{bode_scanning_1997} and hydrogen is considered the most likely species inducing this transition. Recently, presence of hydrogen in Co/Pt(111) films was shown to favor an in-plane orientation of the magnetization easy-axis\cite{ferrer_hydrogen_Co_Pt_2010}. The incorporation of hydrogen into Co/Pd multilayers also modifies their PMA\cite{munbodh_effects_2011}. The origin of these changes has been sometimes adscribed to strain changes due to hydrogen absorption or incorporation to the structure \cite{sander_reversible_2004}, or to unspecified "electronic" effects\cite{munbodh_effects_2011}.

Imaging magnetic domain patterns in the vicinity of SRTs can reveal particularly striking effects that result from inherent frustrations of the system\cite{DeBell_review_2000}. As the magnitude of the anisotropy becomes too small to stabilize a particular easy axis of magnetization, rich pattern-forming phenomena can be observed. Here we describe a new hydrogen-sorption induced SRT that is associated with a ripple-like
fluctuation of the magnetization direction with a wavelength that is consistent with predictions from spin-wave theory\cite{KashubaPRL1993,KashubaPRB1993,BrunoPRB1991,DeBell_review_2000}. 

We start from a thin film system that is known to feature significant dependence of magnetic anisotropy on details of sample structure.
As a function of film thickness, cobalt on Ru(0001) goes through two atomically abrupt consecutive SRTs: one atomically flat layer of cobalt is magnetized in-plane, a bilayer is magnetized out-of-plane, and thicker films are again magnetized in-plane. These transitions can be understood to result from a combination of strain and surface effects\cite{faridprl,gallego_band-filling_2008}. Capping the cobalt films with non-magnetic metal overlayers produces additional SRTs as a function of the thickness of the non-magnetic capping layers\cite{el_gabaly_noble_2008}. All of these SRTs are atomically abrupt, i.e., either in-plane or out-of-plane anisotropy is observed in any given combination of Co and capping layer thickness. 
In the present study, anisotropy in a cobalt bilayer on Ru is changed more gradually by dosing the thin film with controlled, sub-monolayer quantities of hydrogen. Imaging the magnetic domain patterns in close proximity of the SRT, as a function of increasing (or decreasing) hydrogen coverage, permits a real-space observation of the soft spin-wave modes associated with the transition. By means of first-principles calculations we elucidate the physical origin of the experimentally observed SRT. 

\section{Experimental and theoretical details}

The experiments were performed in two low-energy electron microscope (LEEM\cite{altman_trends_2010}) systems. Good base pressures in the 10$^{-11}$ Torr range were maintained in both systems, to permit reasonably good control over hydrogen content of the samples over the time-spans required for measurements. 
The Ru(0001) single crystal substrates are cleaned by flashing to 1700~K in a background pressure of 3$\times 10^{-8}$ Torr of high-purity oxygen. Before cobalt growth, the samples are flashed several times in the absence of oxygen. The cobalt films were grown from electron-bombardment heated deposition sources. The typical flux rate was one atomic Co layer in 3 minutes. Hydrogen was dosed through leak valves from high purity lecture bottles. (To determine the hydrogen dose in Langmuir=$10^{-6}$~Torr$\times$sec, the pressure readings of the ionization gauges were corrected by a factor of 0.46 for hydrogen\cite{gauge} and multiplied by the exposure time). One of the LEEMs, a commercial Elmitec III instrument, was used for diffraction and growth studies. The other instrument uses a spin-polarized electron source and a spin-manipulator in order to permit imaging of the magnetization vector in the Co films. More details on this spin-polarized low-energy electron microscope\cite{grzelakowski_new_1994} (SPLEEM), spin-polarization control\cite{duden_compact_1995} and vector magnetometric application of the instrument can be found in the literature\cite{Ramchal2004PRB,faridprl,el_gabaly_noble_2008,rougemaille_magnetic_2010}.

We have performed {\it ab-initio} calculations within the local density approximation (LDA), combining two different approaches as detailed in Ref.~\onlinecite{gallego_formation_2010}.
First, we have performed an exhaustive search of the equilibrium positions of H using slab models, a plane-wave basis set and the projector-augmented wave (PAW) method \cite{paw}  as implemented in the VASP code \cite{vasp1,vasp1a,vasp2,vasp3}.
For the most stable configurations, we have determined the magnetic anisotropy energy within the fully-relativistic framework of the Screened Korringa-Kohn-Rostoker (SKKR) method \cite{Weinberger2005}. We define the  magnetic anisotropy energy (MAE) according to the convention
\begin{equation}
MAE = E(M_{||}) - E(M_{\perp}),
\end{equation}
where $E$ refers to the total energy, $M_{||}$ to the magnetization in the plane of the surface, and $M_{\perp}$ to perpendicular
magnetization, so that positive values of the MAE correspond to perpendicular magnetic anisotropy (PMA). 
The MAE is obtained based on the magnetic force theorem as the sum of a band energy ($\Delta E_{b}$) term and the magnetostatic dipole-dipole ($\Delta E_{dd}$) term, the second one always favouring $M_{||}$.

\section{Results and Discussion}

\subsection{Experimental results}

We first grow 2~ML thick Co/Ru(0001) films, using elevated growth temperature to favor growth of extended regions with homogeneous thickness [see Fig.~\ref{durante} (a)]. The 550~K substrate temperature used here was selected to optimize film morphology while preventing alloying with the substrate\cite{Farid2007NJPcoru}. Consistent with prior work\cite{faridprl} we find that all 2~ML thick areas are magnetized in the out-of-plane direction [see Fig.~\ref{durante} (b)], while only the 3~ML thick islands show in-plane magnetic contrast [see Fig.~\ref{durante} (g)]. 

This Co film was exposed to molecular hydrogen by filling the chamber to 8$\times$10$^{-10}$ Torr of H$_2$. A sequence of SPLEEM images (out-of-plane electron-beam spin-polarization) was collected in-situ during hydrogen adsorption. Selected frames extracted from the image sequence, reproduced in Fig.~\ref{durante} (b--f), show how the dark and bright out-of-plane magnetic domains break up and disappear as the hydrogen dose is increased. At first, the larger magnetic domains break up into smaller ones forming a disordered stripe pattern, as shown in Fig.~\ref{durante} (b--d).  The area fraction of the out-of-plane domains also begins to decrease. For doses higher than 0.20~L$_{H_2}$ the area fraction decreases more rapidly until all the out-of-plane magnetic contrast disappears, as shown in Fig.~\ref{durante} (d--f). At approximately 0.36~L no out-of-plane magnetic contrast is observed anywhere on the film. 

Repeating these measurements using in-plane polarization of the electron beam confirms that this loss of out-of-plane contrast within the 2~ML thick regions of the film is due to an SRT (and not to loss of magnetization). We briefly heated the film to 400~K and cooled back to room temperature to restore its original, hydrogen-free state (hydrogen is known to desorb from Co at 370~K\cite{bridge_hydrogen_1979}). This is seen in Fig.~\ref{durante} (g): with in-plane polarization of the electron beam, the gray null-contrast in the 2~ML areas arises from their out-of-plane magnetization, while the 3~ML islands present in-plane domains (bright or dark contrast). With hydrogen exposure, in-plane magnetic domains start to be visible in the 2~ML regions of the film, see Fig.~\ref{durante} (h). The domains rapidly evolve with the hydrogen dose and in the neighborhood of 0.24~L$_{H_2}$, Fig.~\ref{durante} (i), we observe again a stripe domain pattern with the same wavevector as before. With increasing hydrogen exposure the magnetic stripe domain phase is replaced with larger in-plane magnetized domains, Fig.~\ref{durante} (j, k), indicating that the SRT is complete.

\begin{figure}[tbh]
\centering \includegraphics[width=0.45\textwidth]{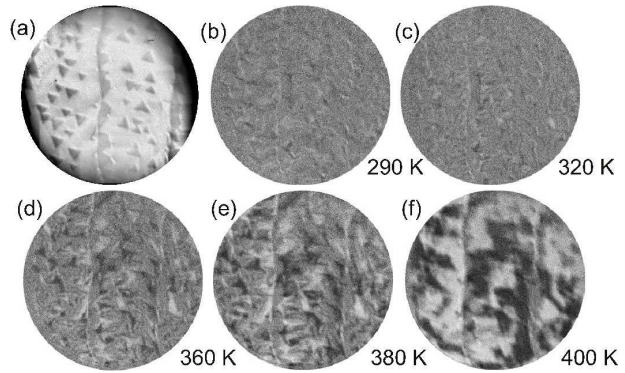}
\caption{(a) LEEM image of a Co film exposed to 0.4~L$_{H_2}$. (b)--(g) Out-of-plane magnetic-contrast SPLEEM images acquired while heating the film shown in (a). The temperature of the sample is shown in each image. FOV is 8~$\mu$m, beam energy 5.2 eV.}
\label{heating}
\end{figure}

The complete reversibility of the H-induced SRT invoked in the previous paragraph was confirmed in more detail. A fresh Co/Ru(0001) film was prepared, cooled down to RT and exposed to 0.4 L$_{H_2}$; a LEEM image of this sample is reproduced in Fig.~\ref{heating} (a). As before, the observation of null-contrast while illuminating this sample with an out-of-plane polarized electron beam indicates in-plane magnetization of the film. As a result of slowly increasing the sample temperature, out-of-plane magnetization of the sample is indeed recovered. This is shown in Fig.~\ref{heating} (b--f). As the sample temperature reaches approximately 300~K [Fig.\ref{heating}(b) and (c)], faint dark and bright features begin to appear, indicating small domains where the out-of-plane component of the magnetization no longer vanishes. As the sample temperature reaches 360~K, Fig.~\ref{heating} (d), the small out-of-plane magnetized domains (bright and dark regions) are resolved more clearly. These domains increase in size until the sample temperature is 400 K, when the magnetic contrast typical of the 2~ML thick Co areas is restored (the Curie temperature of Co bilayers is higher than the hydrogen desorption temperature, Tc = 450~K\cite{faridprl}). The out-of-plane magnetization of this film was observed to persist when cooling down to RT after the annealing step, consistent with the interpretation that the hydrogen-adsorption induced SRT was fully reversed during the thermal desorption step.

To estimate the hydrogen coverage inducing this SRT, we note that the SRT is completed after a total dose of $\sim$0.36~L$_{\mathrm{H_2}}$. By repeating the measurement with the ionization gauge switched off, we confirmed that H dissociation at the hot filament of the ionization gauge does not play a significant role. This dose-measurement also does not depend significantly on the hydrogen pressure: experiments were repeated with hydrogen pressure in the 10$^{-10}$ and 10$^{-9}$ Torr range and the total dose at which out-of-plane magnetic contrast vanishes agreed within 10\%. These observations are consistent with the results from prior thermal programmed desorption studies (TPD, \cite{bridge_hydrogen_1979,christmann_interaction_1988,habermehl-cwirzen_hydrogen_2004}), reporting that hydrogen adsorbs as atomic hydrogen on Co in a non-activated reaction with the reported desorption energy of H/Co in the range of 0.85--1.0~eV/atom\cite{christmann_interaction_1988}. Assuming a constant sticking coefficient of 0.5 up to coverages of 0.5~ML, as reported in Ref.~\onlinecite{lisowski_kinetics_1988}, we estimate that completion of the SRT occurs at a hydrogen coverage in the range of $\Theta$=0.2--0.3.

\begin{figure}
 \centering \includegraphics[width=0.4\textwidth]{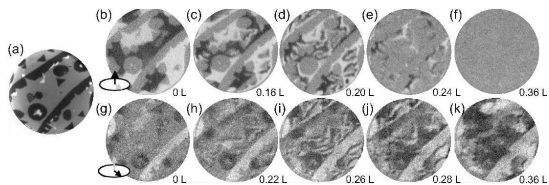}
 \caption{(a)--(b) LEEM images acquired before and after exposing a mostly 2~ML thick Co film on Ru(0001) to molecular hydrogen. FOV is 7$\mu$m, beam energy is 3 eV. (c)--(d) LEED patterns acquired from the central area of the LEEM images, at an beam energy of 72~eV. (e) LEED IV curves acquired from 2 ML Co on Ru(0001) before (continuous curves) and after (dotted curves) hydrogen exposure.}
 \label{doble}
 \end{figure}

Bearing in mind the sensitive dependence of magnetic properties on atomic structure, we used low energy electron diffraction (LEED) measurements to test for possible changes in the structure of Co/Ru films upon hydrogen adsorption. Micro-LEED diffraction patterns acquired from 2 ML thick Co film regions are identical before [Fig.~\ref{doble} (c)] and after hydrogen adsorption [Fig.~\ref{doble} (d)], even after a large dose of 20~L hydrogen was provided. This indicates that the in-plane spacing in the film does not change (within an error of 0.5\%) as a function of hydrogen adsorption. To test for possible hydrogen-induced changes of the interlayer spacing in the films, we measured LEED intensity-versus-voltage spectra. 
Fig.~\ref{doble} (e) shows LEED-IV curves acquired on 2 ML Co/Ru areas before and after the 20~L hydrogen dose. The curves plotted correspond to the spot intensities of the specular beam and two first-order diffracted beams (in all cases, we integrated intensities of main reflections and their satellite spots\cite{Farid2007NJPcoru}). Using Pendry's R-factor\cite{pendry_reliability_1980} to quantify the apparent high degree of similarity of the IV-curves before and after hydrogen adsorption we find an R-factor of 0.04 (mean value for the three beams). This very small value indicates that we have no evidence of any hydrogen-induced change of the interlayer spacing.

\subsection{First-principles calculations}

We have modelled hydrogen covered Co thin films on Ru(0001). The structural changes induced by hydrogen adsorption at different coverages have been discussed in detail in Ref. \onlinecite{gallego_formation_2010}. Here we will focus on 2~ML Co thickness either bare or completely covered by hydrogen. H adsorbs preferentially at hollow positions, with a slight preference (around 25 meV)
for fcc stacking over hcp. Other high symmetry adsorption sites (bridge, ontop) are much less favorable
(over 200 meV for the most favorable bridge sites) and subsurface positions are less stable by more than 
400 meV. These energy barriers strongly limit the possibility of H diffusion either across the surface or
into the bulk.

In agreement with the experimental results, H hardly modifies the Co/Ru structure: the variation of interlayer distances upon H adsorption remains below 0.07~\AA, and reflects mainly an attenuation of the surface induced effects. Nevertheless, the H-Co bond is strong, and the H covered surfaces are much more stable than the bare ones, as evidenced by the calculated work functions and adsorption energies\cite{gallego_formation_2010}.

\begin{figure}[th]
\begin{center}
\includegraphics[width=\columnwidth,clip]{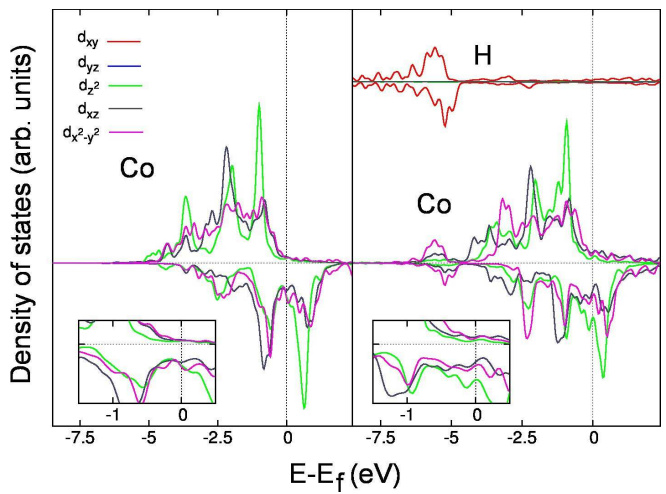}
\caption{ (Color online)
Spin-resolved DOS without SOC projected on the H and outermost Co atoms for the (left) bare and (right) H covered surfaces. 
Positive (negative) DOS values correspond to majority (minority) spin polarizations. Only the $s$ states of H and
the $l,m$ resolved $d$ band of Co are shown. The insets show an amplified image of the corresponding DOS of Co around E$_F$.}
\label{fdos}
\end{center}
\end{figure}

Even though the structural changes are minor, significant H signatures can be found in the electronic structure
(see Fig.~\ref{fdos}). H states are located at the bottom of the valence band, with strong hybridization to the outermost
Co atoms, and have a low influence on those Co not bonded to H (either subsurface Co or surface Co atoms for partial H coverage). The effect of H is to reduce the Co magnetic moments (see table III of Ref.~\onlinecite{gallego_formation_2010}),
due to the broadening of the density of states (DOS) and the reduction of the on-site exchange. 
This alters significantly the partial occupation of $d$ bands, as shown in Fig.~\ref{fdos}.
It is evident from the figure that the presence of H modifies the occupation of the states around the Fermi
level (E$_F$), increasing the total DOS and the corresponding partial occupancies of all $m$ projections, specially
$d_{z^2}$. As we will show below, this effect determines the magnetic anisotropy.

Table \ref{tlayeb} shows the layer-resolved contribution to the band energy both for the bare and H covered surfaces. 
Calculations with H at the hcp adsorption positions do not modify the results qualitatively, though a much moderate value of $\Delta$E$_b$ for the surface Co layer is obtained (-0.142 eV).
H suppresses the positive band energy favoring PMA, not only changing the sign of $\Delta$E$_b$, but also strongly favoring an in-plane orientation of the magnetization. The effect is local, only affecting the outermost Co atoms bonded to H; this has been confirmed
in calculations for thicker Co films. However, these surface atoms carry the highest contribution to PMA at the bare
surface, required to overcome the dipolar term $\Delta$E$_{dd}$ of about 0.16 eV. Thus, the presence of hydrogen leads to an in-plane orientation of the magnetization.

\begin{table}[htbp]
\caption{Layer resolved and total contribution to $\Delta$E$_b$, in eV, for 2 Co/Ru(0001) depending on the H coverage.
Ru,Co refer to the atoms at the interface, Co$_s$ to the surface Co atom.}
\begin{tabular}{cccccc}
\hline \hline
   &  Ru  &  Co    & Co$_s$ &  H      & Tot \\
\hline
 Bare & -0.032 &  0.111 &  0.116 &   --   & 0.188 \\
 1 ML & -0.022 &  0.128 & -0.294 & -0.029 &  -0.228\\
\hline \hline
\end{tabular}
\label{tlayeb}
\end{table}

\begin{figure}[th]
\begin{center}
\includegraphics[width=\columnwidth,clip]{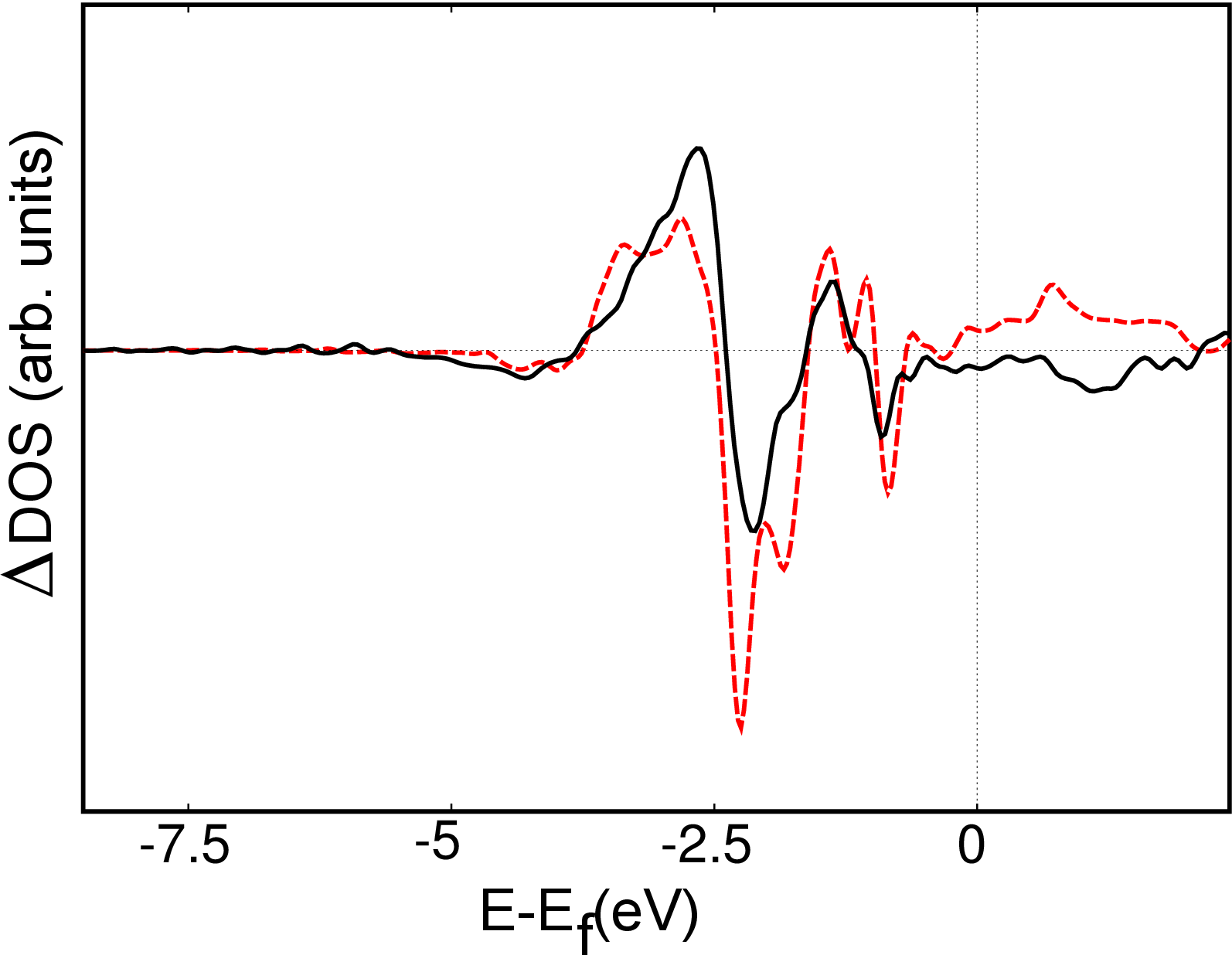}
\caption{ (Color online)
Cumulative difference of the DOS corresponding to a perpendicular and an in-plane orientation of the magnetization
projected on the $d$ band of the outermost surface Co atom for a (dashed red) bare and
(solid black) H covered surface. The y-scale is 10 times smaller than in Fig.~\protect\ref{fdos}.}
\label{fcumdos}
\end{center}
\end{figure}

The strong mixing of hybridized states of different orbital character makes it difficult to assign the effect to any 
particular $m$ projection, specially taking into account that the introduction of the spin-orbit coupling (SOC) 
breaks the degeneracy of $\pm m$ states, altering both their splitting and ordering depending on the orientation 
of the magnetization with respect to the lattice. However, the crucial effect of H can be clearly identified regarding 
the difference between the $d$ DOS projected on the Co surface atoms for an in-plane and a perpendicular
orientation of the magnetization. This cumulative energy-resolved difference is plotted in Fig.~\ref{fcumdos}.
According to our convention and within the magnetic force theorem, the change of sign of the cumulative DOS in the
presence of H corresponds to that of the MAE \cite{lehnertPRB2010}. The important changes occur in a narrow energy window of around 0.5 eV
below E$_F$. Comparing to the DOS at Fig.~\ref{fdos}, the major changes in this energy range upon H load are the 
shift of the minority spin DOS to higher occupancies and an increased contribution of $d_{z^2}$ states. 

To summarize, H adsorption produces a spin reorientation transition in ultrathin Co films on Ru(0001),
by strongly favoring an in-plane orientation of the magnetization. The effect is purely electronic, ultimately rooted
in the strong Co-H binding and the subsequent modification of the Co electronic states around E$_F$.

\subsection{Discussion}

Our experimental and theoretical results clearly indicate that hydrogen adsorption on the cobalt bilayer on Ru(0001) produces a spin-reorientation transition, changing the magnetization from an out-of-plane orientation to in-plane. Its origin can be explained from the ab-initio calculations by the hybridization of the hydrogen orbital and the orbitals of the Co atoms bonded to the absorbed hydrogen. It is thus a local electronic effect rather that a strain-based effect such as the one detected in Ni films\cite{sander_reversible_2004}.  

Before the SRT occurs, the out-of-plane magnetic domains are observed to break-up into smaller domains forming a disordered stripe pattern. In films with perpendicular magnetic anisotropy, as a SRT is approached, the appearance of stripe magnetic domain patterns has been observed in a number of systems\cite{won_magnetic_2005,choi_magnetic_2007,wu_magnetic_2004,pescia_fecustripes_2010}. 
Our SPLEEM images show that in the neighborhood of 0.24~L$_{\mathrm{H_2}}$ dose, the width of the out-of-plane magnetized stripe domains [Fig.~\ref{durante} (d)] is approximately equal to the width of the in-plane magnetized regions [Fig.~\ref{durante} (i)], and of the order of 0.24$\pm$0.05 $\mu$m. This number is in the range of experimental results of other quasi two-dimensional cobalt films, such as the experimental results by Won et al. [0.2--0.4 $\mu$m for Fe/Ni/Cu(100)]\cite{wu_magnetic_2004,won_magnetic_2005}.
The domain size is the result  of the balance between the magnetic anisotropy, and the exchange and dipolar interactions\cite{KashubaPRL1993,KashubaPRB1993,BrunoPRB1991,DeBell_review_2000}. While in many studies only the first order magnetic anisotropy is considered, second order anisotropies can be substantial and may affect strongly the SRT, such as whether the magnetization direction changes continuously or discontinuously across the transition\cite{Gradmann1994,Kirschner1997}. In the hydrogen on cobalt experiments presented here, the sequential character of the observations (first in-plane, then out-of-plane in consecutive transitions) prevents us from a detailed statement on the character of the transition. Nevertheless,
a ripplelike instability of the homogeneously magnetized state is expected close to the SRT,
due to the competition between magnetic dipolar interactions and magnetocrystalline anisotropy \cite{BrunoPRB1991}.

This competition is also reflected in our calculations. Several factors act simultaneously when exposing the cobalt films to hydrogen. The average $\Delta$E$_{b}$ of the film is expected to gradually decrease, and eventually change sign, as the coverage of hydrogen is increased. While this modification is probably the main effect responsible for the SRT, the magnetic moments of the surface Co atoms also depend on hydrogen coverage. The dipolar magnetic anisotropy energy is, thus, also modulated by hydrogen. Furthermore, this evolution in magnetic moment is not linear with coverage\cite{gallego_formation_2010}.

One final detail is that the observed SRT does not occur homogeneously over the film (see, for example, Fig.~\ref{durante}e). In principle, the SRT should happen when a threshold local hydrogen concentration is reached. To explain the non-uniform SRT, the hydrogen concentration should in turn be non-uniform. This can be due to either defects or other adsorbed gases on the surface of the film, limited surface diffusion, or a bidimensional phase transition on the adsorbed hydrogen. Further work will be needed to determine the precise origin of this effect.

\section{Summary}

In summary, by means of SPLEEM we have discovered that room-temperature adsorption of a dose of $\sim$0.36~L$_{\mathrm{H_2}}$ induces a spin-reorientation transition from out-of-plane to in-plane on 2~ML thick Co regions on Ru(0001). This SRT is reversed by heating the films to 400 K, a consequence of thermal desorption of the hydrogen. By means of first-principle calculations, we determine that the origin of the change in the magnetic anisotropy energy of the film is the hybridization of the hydrogen and Co atoms closest to the adsorbed hydrogen. The effect is nearly a pure electronic one, in contrast to other systems where it is due to the emergence of strain in the metal film. The SRT occurs non-uniformly on the film, indicating a non-uniform hydrogen concentration. Before the SRT, the domain size decreases and reaches a limiting value of 0.24~$\mu$m.

It is possible that the sensitivity of magnetic anisotropy to hydrogen adsorption is a more common phenomenon that is not restricted to this case of 2~ML Co/Ru and the small number of other systems described in the literature\cite{sander_reversible_2004, ferrer_hydrogen_Co_Pt_2010}. Recalling that hydrogen is the most common component in the residual gas of most ultra-high vacuum chambers, these observations remind us of the basic challenges of surface science experimentation. It is also conceivable that the high sensitivity of the Co magnetization easy-axis to small doses of hydrogen could be employed in devices designed to detect and signal the presence of hydrogen\cite{sensor}. 

\begin{acknowledgments} 
This research was supported by the Spanish Ministry of Science and Technology through
Projects No. MAT2009-14578-C03-01 and MAT2010-21156-C03-02 and the Office of Basic Energy Sciences, Divisions of Materials and Engineering Sciences, U. S. Department of Energy under Contracts No. DE-AC04-94AL85000 and DE-AC02-05CH11231. BS thanks the Spanish Ministry of Science and Innovation for support through an FPI fellowship and ATN thanks the Alexander von Humboldt Foundation for financial support.
\end{acknowledgments}

%

\end{document}